# EXPERIMENTAL AND MODELING STUDY OF THE AUTOIGNITION OF CYCLOPENTENE

M. YAHYAOUI, M. H. HAKKA, GLAUDE P.A. and BATTIN-LECLERC F.

*Département de Chimie-Physique des Réactions, UMR 7630 CNRS-INPL, Nancy Université*

*1, rue Grandville - BP 20451 - 54001 NANCY Cedex - France*

## ABSTRACT

Ignition delay times of cyclopentene-oxygen-argon mixtures were measured behind reflected shock waves. Mixtures contained 0.5 or 1 % of hydrocarbons for equivalence ratios ranging from 0.5 to 1.5. Reflected shock waves conditions were: temperatures from 1300 to 1700 K and pressures from 7 to 9 atm. When comparing to previous results obtained under similar conditions, it can be observed that the reactivity of cyclopentene is much lower than that of cyclohexene, but very close to that of cyclopentane. A kinetic mechanism recently proposed for the combustion of cyclopentene in a flame has been used to model these results and a satisfactory agreement has been obtained. The main reaction pathways have been derived from flow rate, simulated temporal profiles of products and sensitivity analyses.

## INTRODUCTION

The formation of PAHs and soot represents one of the most intriguing problems for kineticists and questions still remain even concerning the formation and the oxidation of the first aromatic compounds. While it is now well accepted that the formation of the first aromatic ring occur in most cases through the recombination of propargyl radicals [1-3], recent work has suggested a potential link between $C_6$ and $C_5$ cyclic species [4,5] and the self-recombination of cyclopentadienyl radicals is a possible way of formation of naphthalene [6]. The oxidation of cyclopentene has already been experimentally investigated in a shock tube [7] and in laminar premixed flames with cyclopentene as only fuel [8, 9]. Lindstedt and Rizos [10] have proposed a model to simulate the results of Lamprecht et al. [8]. We have also recently experimentally analysed and modeled the structure of a laminar premixed methane flame doped with cyclopentene [11] and the purpose of the present work is to extend the range of validity of our model by studying autoignition delay times of cyclopentene in a shock tube.

## EXPERIMENTAL

The shock tube measurement of ignition delay times has been described in several papers [12-18]; the main features of this experimental device will just be recalled here. Ignition delay times have been measured in a stainless steel shock tube; the reaction and the driver parts were respectively 400.6 and 89 cm in length and were separated by two terphane diaphragms. Both sections of the shock tube were evacuated using two primary vacuum pumps. The diaphragms were ruptured by decreasing suddenly the pressure in the space separating them, that allowed to keep the same pressure in the high pressure part for all experiments. The driver gas was helium. The shock velocity was measured via four pressure transducers equally spaced by 150 mm, mounted flush with the inner surface of the tube, the last one being 2 mm before the shock tube end wall. Reflected shock conditions ($P_5$, $T_5$) were calculated from standard procedure, using energy and momentum conservation. A fused silica window (9 mm optical diameter and 6 mm thickness) was mounted across a monochromator centred at 306 nm, which is characteristic of OH chemiluminescence emission, and equipped with an UV-sensitive photomultiplier tube. The window was located at the same place along the axis of the tube as the last pressure transducer. The time response of the emission detection device is around 3 µs.

FIGURE 1

Figure 1 showed an example of the recorded signals, representing pressure evolution and OH emission. It is worth noting that the experimental OH emission at 306 nm is related to electronically exited OH*. The ignition delay time was defined as the time interval between the pressure rise measured by the last pressure transducer due to the arrival of the reflected shock wave and the rise of the optical signal delivered by the photomultiplier up to 50% of its maximum value, as represented as $\tau_{ign}$ on the Figure 1. The small increase in the pressure between the reflected shock wave and the auto-ignition is due to the heat release during the first stage of the oxidation. The larger increase after the ignition may be due to a backward reflection of the shock wave on the surface between driver and reaction gases. The experiment is valid only if the ignition occurs before this reflection. Electronic time responses are short compared to the ignition delay times but some uncertainties comes from the determination of the ignition time at 50% of the OH* radical peak height. The temperature range for the investigation is limited by the ignition delay times that are kept between 10 µs and 1000 µs to avoid too large uncertainties.

Cyclopentene was purchased from Fluka, with a purity of 99% and was degassed several times before the mixtures were prepared. Oxygen, argon and helium were purchased from Messer. Fresh reaction mixtures were prepared every day in a 20L tank and mixed using a recirculation pump. The blend was prepared by adding the partial pressure of each gas to reach a total pressure of 800 Torr. Before each introduction of the reaction mixture, the reaction section was flushed with pure argon and evacuated, for insuring the residual gas to be mainly argon.

This study was performed under the following experimental conditions, after the reflected shock:

- Temperature range from 1300 to 1700 K,

- Pressure maintained around 8.5 atm, ranging from 7.5 to 9.5 atm,

- Mixtures (argon / cyclopentene / oxygen, in molar percent) were (92.5 / 0.5 / 7), (96 / 0.5 / 3.5), (97.2 / 0.5 / 2.3) and (92 / 1 / 7), respectively, corresponding to three different equivalence ratios ($\varphi$ = 0.5, 1 and 1.5) and to two different concentrations of cyclopentene (0.5 and 1 %) and allowing to obtain delay times from 15 up to 1270 µs.

**EXPERIMENTAL RESULTS**

Table I summarizes the experimental measurements performed in this study. Figure 2 presents the experimental results obtained for the ignition of cyclopentene in the case of the mixture containing 0.5% of hydrocarbon, for three equivalence ratios. Figure 3 displays the experimental results obtained for hydrocarbon concentrations of 1% and 0.5% with a constant equivalence ratio of 1. These results show that in each case, ignition delay times increase when equivalence ratio increases and decrease when dilution decreases.

<div align="center">

**TABLE I**

**FIGURES 2 & 3**

</div>

For hydrocarbon / oxygen mixtures, the determination of power dependences is often proposed from the overall statistical correlation between $\tau$ and the gas concentrations:

$$t_{igni} = A \exp(E/RT) [HC]^a [O_2]^b [Ar]^c$$

where A is the pre-exponential factor and E the apparent activation energy. For a restricted range of pressure and temperature, a, b and c are usually constant. Such a statistical correlation has been derived from the present experiments, but since the mole fraction of argon had small variations under the different initial conditions (from 92 to 97.2%) and seems not to affect the delay times, it was chosen to keep c = 0. A multi-linear regression gave the following relationship, with the concentrations behind the reflected shock wave in mol.cm$^{-3}$:

$$t_{igni} (s) = 8.95 \times 10^{-18} \exp(56470/RT) [cyclopentene]^{0.73} [O_2]^{-1.73}$$

The statistical correlation shows a strong negative $O_2$ power dependence, while the fuel power dependence is near 0.7. Figure 4 compares this correlation with experimental data by plotting the ratio between experimental delay times and components concentrations versus $1/T_5$. The spread is linear and shows a good agreement of the global "Arrhenius" behavior of the relationship with the experimental data.

**FIGURE 4**

This correlation can be compared with that proposed by Burcat et al. [7] for a similar study over a concentration range of 0.25 to 1% cyclopentene and 1,75-7 % oxygen for a temperature range of 1323-1816 K and a pressure range 1.67-7.36 atm:

$$t_{igni} (s) = 8.95 \times 10^{-16} \exp(52770/RT) [cyclopentene]^{0.59} [O_2]^{-1.61} [Ar]^{0.33}$$

It can be observed that the activation energy is close to that obtained in the present study, as well as the order for cyclopentene and oxygen. Figure 3 displays a plot of the correlation of Burcat et al. [7] for our experimental conditions corresponding to an equivalence ratio of 1 and 1% cyclopentene and shows that there is a very good agreement with our measurements.

Figure 5 presents a comparison between ignition delay times of cyclopentene, cyclopentane [17], cyclohexene [16] and linear 1-pentene [18] obtained under the same conditions, i.e. a fuel concentration of 1% and an equivalence ratio of 1. The pressure range was the same for all the studies, from 7.3 to 9.5 atm with a mean value of 8.4 atm. Consistent with previous findings, the reactivity of cyclopentane is much lower than that of cyclohexane, we can observe here that the reactivity of cyclopentene is much lower than that of cyclohexene, but very close to that of cyclopentane. Figure 5 also shows that the reactivity of cyclopentene is much lower than that of 1-pentene.

**FIGURE 5**

**KINETIC MECHANISM**

The mechanism that we have used here is based on a previous one recently developed to model the combustion of cyclopentene, as an additive in a laminar premixed methane flame at low pressure [11]. It includes previous the mechanisms that were built to model the oxidation of $C_3$-$C_4$ unsaturated hydrocarbons [12, 13, 19, 20], benzene [14] and toluene [15]. Thermochemical data are estimated by the software THERGAS developed in our laboratory [21], which is based on the additivity methods proposed by Benson [22].

*Reaction base for the oxidation of C3-C4 unsaturated hydrocarbons*

This C3-C4 reaction base, which is described in details in the previous papers [12, 13, 19, 20], was built from a review of the recent literature and is an extension of our previous $C_0$-$C_2$ reaction base [23]. The $C_3$-$C_4$ reaction base includes reactions involving $C_3H_2$, $C_3H_3$, $C_3H_4$ (allene and propyne), $C_3H_5$, $C_3H_6$, $C_4H_2$, $C_4H_3$, $C_4H_4$, $C_4H_5$, $C_4H_6$ (1,3-butadiene, 1,2-butadiene, methyl-cyclopropene, 1-butyne and 2-butyne), $C_4H_7$ (6 isomers), as well as some reactions for linear and branched $C_5$ compounds and the formation of benzene. In these reactions bases, pressure-dependent rate constants follow the formalism proposed by Troe [24] and efficiency coefficients have been included.

*Mechanisms for the oxidation of benzene and toluene*

Our mechanism for the oxidation of benzene contains 135 reactions and includes the reactions of benzene and of cyclohexadienyl, phenyl, phenylperoxy, phenoxy, hydroxyphenoxy, cyclopentadienyl, cyclopentadienoxy and hydroxycyclopentadienyl free radicals, as well as the reactions of ortho-benzoquinone, phenol, cyclopentadiene, cyclopentadienone and vinylketene, which are the primary products yielded [14].

The mechanism for the oxidation of toluene contains 193 reactions and includes the reactions of toluene and of benzyl, tolyl, peroxybenzyl (methylphenyl), alcoxybenzyl and cresoxy free radicals, as well as the reactions of benzaldehyde, benzyl hydroperoxyde, cresol, benzylalcohol, ethylbenzene, styrene and bibenzyl [15]

*New mechanism proposed for the oxidation of cyclopentene*

As this mechanism is described in details in a previous paper [11], we recall here only its main features. We have considered the unimolecular reactions of cyclopentene, the additions to the double bond of H-atoms and OH radicals and the H-abstractions by oxygen molecules and small radicals. Unimolecular reactions include dehydrogenation to give cyclopentadiene, decompositions by breaking of a C-H bond and isomerization to give 1,2-pentadiene.

The reactions of cyclopentenyl radicals involved isomerizations, decompositions by breaking of a C-C bond to form linear $C_5$ radicals including two double bonds or a triple bond, the formation of cyclopentadiene by breaking of a C-H bond or by oxidation with oxygen molecules and terminations steps. Termination steps were written only for the resonance stabilized cyclopentenyl radicals: disproportionnations with H-atoms and OH radicals gave cyclopentadiene, combinations with $HO_2$ radicals led to ethylene and $CH_2CHCO$ and OH radicals, and combinations with $CH_3$ radicals formed methylcyclopentene.

The decomposition by breaking of a C-H bond of cyclopentyl radicals led to the formation of 1-penten-5-yl radicals. The isomerizations (for the radical stabilized isomer) and the decompositions by breaking of a C-C bond of the linear $C_5$ radicals were also written, while those by breaking of a C-H bond were not considered.

The reactions of cyclopentadiene were part of the mechanism for the oxidation of benzene, but reactions for the consumption of methylcyclopentene and methylcyclopentadiene, obtained by recombination of cyclopentadienyl and methyl radicals had to be added.

The kinetic data were not modified compared to our previous paper [11], except for the addition of H-atoms to cyclopentene to give cyclopentyl radicals. We have used here the high pressure limit value calculated by Sirjean et al. [17] by theoretical methods.

**DISCUSSION**

Simulations have been run by using the SENKIN code of the Chemkin II [25] software library. The simulations have been performed for an average pressure of 8.2 atm, considering an adiabatic reactor with a constant volume. Simulated ignition delay times have been taken as the time at 50 % of the maximum concentration of excited OH* radicals using the mechanism for excited species developed by Hall et al. [26]. Figures 2 and 3 present the comparisons between computed and experimental results in the cases of concentration of respectively 0.5 and 1% of cyclopentene. The agreement obtained is rather satisfactory.

Figure 6 shows the scheme of the significant reactions at 1437 K for a stoichiometric mixture and for a calculated conversion of 20%. Figure 7 presents the temporal profiles of cyclopentene and the main carbon containing products obtained. This figure also displays the time evolution of OH and OH* radicals and shows that the delay times derived from the rise of both radicals are very close. Figure 8 presents a sensitivity analysis performed under the conditions of fig.6 at 1437 and 1703 K. For each studied reaction, ignition delay time has been computed with a mechanism in which this given reaction has been removed and compared with the result obtained with the full mechanism and with a mechanism in which the rate constant of the very accelerating reaction (H+$O_2$=O+OH) has been multiplied by 10. This last reaction is a branching step and all our previous work in the field of ignition in a shock tube has shown that its rate constant is by far the most sensitive parameter [12-18].

**FIGURES 6, 7, 8**

Under the conditions of figure 6, cyclopentene is mainly consumed by molecular dehydrogenation to give hydrogen and cyclopentadiene (reaction R1 in figure 6, 58% of the consumption of cyclopentene at 1437 K, 37% at 1707 K), by addition of H atoms to form cyclopentyl radicals (R3, 27% of the consumption at 1437 K, 35% at 1707 K) and by unimolecular decomposition to produce stabilized cyclopentenyl radicals (R5, 9.5% of the consumption at 1437 K, 20% at 1707 K). Reaction R1 has an inhibiting effect, while competing reactions R3 and R5 have a promoting effect. Autoignition occurs far later after the complete consumption of cyclopentene and seems more triggered by the complete consumption of cyclopentadiene as shown in fig. 7.

Minor reactions include H-abstraction, mainly by H atoms and OH radicals, to produce resonance stabilized cyclopentenyl radicals (R6, 3.2% of the consumption at 1437 K, 2.5% at 1707 K), alkenyl cyclopentenyl radicals (R7, 1.2% of the consumption at 1437 K, 0.6% at 1707 K) and vinylic cyclopentenyl radicals (R4, 0.07% of the consumption at 1437 K, 0.6% at 1707 K) and isomerization to give 1,2-pentadiene and then 1,3-pentadiene (reaction R2, 0.15% of the consumption at 1437 K, 2% at 1707 K).

Cyclopentyl radicals isomerize rapidly to form 1-penten-5yl radicals. These linear radicals are almost completely decomposed to give ethylene and resonance stabilized allyl radicals. The main reaction of these last radicals is isomerisation to give $sC_3H_5$ radicals, which readily decompose to give methyl radicals and acetylene. Cyclopentadiene, ethylene and acetylene are the major reaction products as shown in figure 7. The combination of allyl radicals with methyl radicals leads to 1-butene and their beta-scission by H-elimination produces allene. A very small fraction of 1-penten-5yl radicals produces 1,3-pentadiene by H-elimination. Allene, 1-butene and 1,3-pentadiene are also important products of the reaction.

The exclusive fate of allylic and alkenyl cyclopentenyl radicals is to form cyclopentadiene and H-atoms. Vinylic cyclopentenyl radicals react by the opening of the cycle leading ultimately to ethylene and propargyl radicals which are a source of benzene under these high pressures.

Under the conditions of fig. 6, cyclopentadiene is consumed to produce cyclopentadienyl radicals either by H-abstractions (reaction R8) or by unimolecular initiation (reaction R9). H-abstractions transform reactive H atoms to give resonance stabilized cyclopentadienyl radicals and have therefore a strong inhibiting effect. The concurrent unimolecular initiation, which produces H-atoms, has consequently a promoting effect, as shown in fig. 8a at 1703K. A flow analysis, at 1347 K just before ignition, shows the unimolecular initiation does not occur anymore, but that cyclopentadienyl radicals combine with H-atoms explaining the inhibiting effect of reaction R9 at this low temperature.

Cyclopentadienyl radicals are consumed by reaction with oxygen molecules to give $C_4H_4O$ and HCO radicals (reaction R12), by combination with methyl (reaction R10) and $HO_2$ (reaction R13) radicals, by self-combination (reaction R11) and by addition to acetylene to produce benzyl radicals (reaction R14). The rate of consumption of these resonance stabilized radicals is almost 10 times lower than their rate of formation and is then a limiting factor. That explains why reactions R11 and R12, which involves the formation of reactive H atoms or OH radicals, have nevertheless an inhibiting effect, i.e. the removal of these reactions reduces the flow rate of formation of cyclopentadienyl radicals and decreases then delay times. Combinations with methyl radicals lead to methylcyclopentadiene, those with HO2 radicals to cyclopentadienone and self-combination to bicyclopentadienyl. These three compounds are important reaction products as shown in fig. 7.

Aromatic species are also reaction products, present in noticeable amounts. There are two equivalent sources of benzene: the self-combination of propargyl radicals and the isomerization of cyclopentadienyl methylene radicals, which are obtained either directly by H-abstractions by H-atoms or OH radicals from methylcyclopentadiene or by isomerisation from the resonance stabilized methyl cyclopentadienyl radicals, which are also obtained by H-abstractions by H-atoms or OH radicals from methylcyclopentadiene. Contrary to the case of our low-pressure flame (6.7 kPa) [11], under the conditions studied here, the combination of propargyl radicals leads directly to benzene and not to phenyl radicals, which can easily consumed by oxygen molecules. Toluene is obtained from benzyl radicals by H-abstraction from cyclopentadiene or combination with H-atoms.

According to flow rate analyses at 1437 K, about 70% of the consumption of cyclopentene involves the consumption of reactive radicals, e.g. H atoms or OH radicals, to give resonance stabilized radicals, such as cyclopentadienyl radicals by H-abstraction from cyclopentene or allyl radical by H-addition to cyclopentene. The oxidation of cyclopentane has the same characteristics: under the same conditions, about 70% of their consumption involves the transformation of reactive radical into resonance stabilized allyl radicals, by H-abstraction from cyclopentane followed by the decomposition of cyclopentyl radicals [17]. That explains why both $C_5$ cyclic compounds have a close reactivity. In the case of cyclohexene under similar conditions, the formation of resonance stabilized radicals is less favored and cyclohexadienyl radicals can easily decompose to give benzene and H-atoms, which promotes ignition [16].

The occurrence of unimolecular initiations from cyclopentene and cyclopentadiene involving the breaking of a C-H bond are favored by the formation of resonance stabilized radicals. This type of reactions is of much lower importance from cyclopentane and explains why simulations predict a slightly higher reactivity of the unsaturated species at low temperatures. In the case of 1-pentene [18], unimolecular initiations involving the breaking of a C-C bond (about 40% of the consumption of 1-pentene) rapidly leads to the formation of H-atoms and strongly promotes the reactivity. In the case of cyclic species, unimolecular initiations involving the breaking of a C-C bond lead to diradicals, which further react to give linear molecules which can then decompose to give radicals, with only a slight promoting effect on the reactivity [17].

**CONCLUSION**

This paper presents new measurements concerning ignition delay times of cyclopentene in a shock tube behind reflected shock wave from T = 1300 - 1700 K, as well as simulations based on a previously develop detailed kinetic model [11] and allowing us to correctly reproduce these results. Experimental results show that cyclopentene has a reactivity close to that of cyclopentane and much lower than that of cyclohexene and that of 1-pentene. Simulations show that the main reaction products are cyclopentadiene, ethylene and acetylene. Most of the reaction pathways involve the transformation of reactive radicals, e.g. H atoms or OH radicals, into resonance stabilized radicals, such as cyclopentadienyl radicals or allyl radical, which explains the low reactivity of this compound.

**LEGENDS OF FIGURES**

Figure 1: Evolution of the pressure and OH emission signals recorded as a function of the time.

Figure 2: Ignition delay times of cyclopentene in shock tube for equivalence ratios of 0.5, 1 and 1.5 and a concentration of hydrocarbon of 0.5%. Points correspond to experimental results and lines to simulations performed for a constant pressure of 8.2 atm.

Figure 3: Ignition delay times of cyclopentene in shock tube for an equivalence ratio of 1 and concentrations of hydrocarbon of 1% and 0.5%. Points correspond to experimental results and lines to simulations, performed for a constant pressure of 8.2 atm, or to the correlation of Burcat et al. [7].

Figure 4: Comparison of experimental results and statistical Arrhenius fit vs. temperature for every condition studied. Points correspond to experimental results and the line is the best-fit to experimental data.

Figure 5: Comparison between ignition delay times of cyclopentene, cyclopentane [17], cyclohexene [16] and 1-pentene [18] for an equivalence ratio of 1 and a concentration of hydrocarbon of 1%. Points correspond to experimental results and lines to simulation.

Figure 6: Major reaction pathways in the case of the stoichiometric oxidation of cyclopentene (initial concentration of hydrocarbon of 0.5%) at a temperature of 1437 K and a pressure of 8.2 atm. Simulated conversion is 20%. The temporal profiles of the products in bold are shown on figure 7.

Figure 7: Simulated temporal profiles of the consumption of cyclopentene and the formation of the main products before autoignition under the conditions of fig.6.

Figure 8: Sensitivity analysis for the main reactions of cyclopentene, cyclopentadiene and cyclopentadienyl radicals under the conditions of fig.6 at (a) 1437 K and (b) 1703 K

(Rx refer to the reactions shown in fig. 6.; for metatheses, all the reactions have been removed regardless of the radical involved un the reaction).

**TABLE I: Mixture compositions, shock conditions and ignition delay times for cyclopentene.**

| Composition (mole Percent) | | $P_5$ | $T_5$ | $\tau$ | Composition (mole Percent) | | $P_5$ | $T_5$ | $\tau$ |
|---|---|---|---|---|---|---|---|---|---|
| $C_5H_8$ | $O_2$ | (kPa) | (K) | (µs) | $C_5H_8$ | $O_2$ | (kPa) | (K) | (µs) |
| 0.5 | 3.5 | 820 | 1437 | 355 | 0.5 | 2.3 | 853 | 1504 | 281 |
| | | 820 | 1465 | 245 | | | 892 | 1538 | 193 |
| | | 823 | 1495 | 237 | | | 868 | 1571 | 125 |
| $\varphi = 1$ | | 840 | 1524 | 93 | $\varphi = 1.5$ | | 765 | 1620 | 66 |
| | | 835 | 1539 | 126 | | | 786 | 1628 | 71 |
| | | 850 | 1554 | 86 | | | 821 | 1667 | 46 |
| | | 817 | 1561 | 91 | | | 852 | 1702 | 34 |
| | | 822 | 1604 | 72 | | | 800 | 1720 | 33 |
| | | 780 | 1609 | 78 | | | 853 | 1750 | 32 |
| | | 895 | 1649 | 51 | | | 833 | 1759 | 21 |
| | | 832 | 1702 | 15 | | | 842 | 1772 | 18 |
| | | | | | | | 843 | 1773 | 15 |
| 0.5 | 7 | 866 | 1344 | 447 | 1 | 7 | 836 | 1311 | 953 |
| | | **847** | **1360** | **294** | | | 807 | 1317 | 1268 |
| | | 878 | 1367 | 217 | | | 865 | 1339 | 973 |
| $\varphi = 0.5$ | | 873 | 1376 | 191 | $\varphi = 1$ | | 777 | 1346 | 814 |
| | | 848 | 1380 | 203 | | | 837 | 1368 | 275.8 |
| | | 863 | 1404 | 220 | | | 798 | 1370 | 470 |
| | | 936 | 1428 | 132 | | | 763 | 1388 | 327 |
| | | 877 | 1449 | 83 | | | 839 | 1410 | 246.7 |
| | | 781 | 1468 | 46 | | | 824 | 1414 | 191.7 |
| | | 818 | 1478 | 43 | | | 857 | 1414 | 214.2 |
| | | 837 | 1507 | 38 | | | 850 | 1456 | 112 |
| | | 814 | 1532 | 30 | | | 842 | 1479 | 76 |
| | | 829 | 1550 | 30 | | | 826 | 1530 | 46 |
| | | | | | | | 768 | 1585 | 30 |
| | | | | | | | 839 | 1592 | 38 |
| | | | | | | | 855 | 1641 | 20.3 |

*Note*: $P_5$ and $T_5$ are pressure and temperature behind the reflected shock wave, $\tau$ is the ignition delay time. The data in bold correspond to the conditions of figure 1.

**FIG 1**

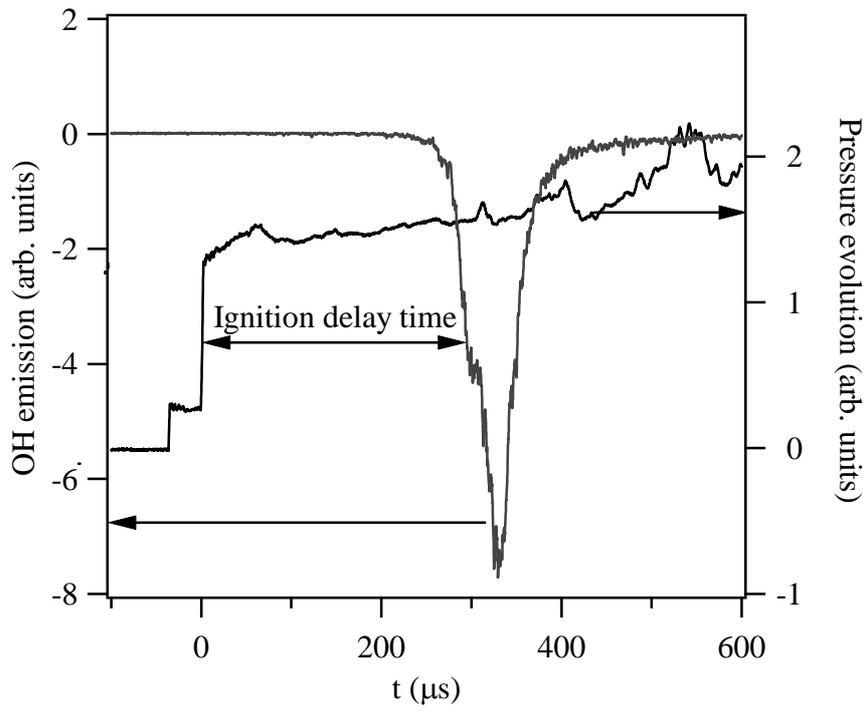

**FIG 2**

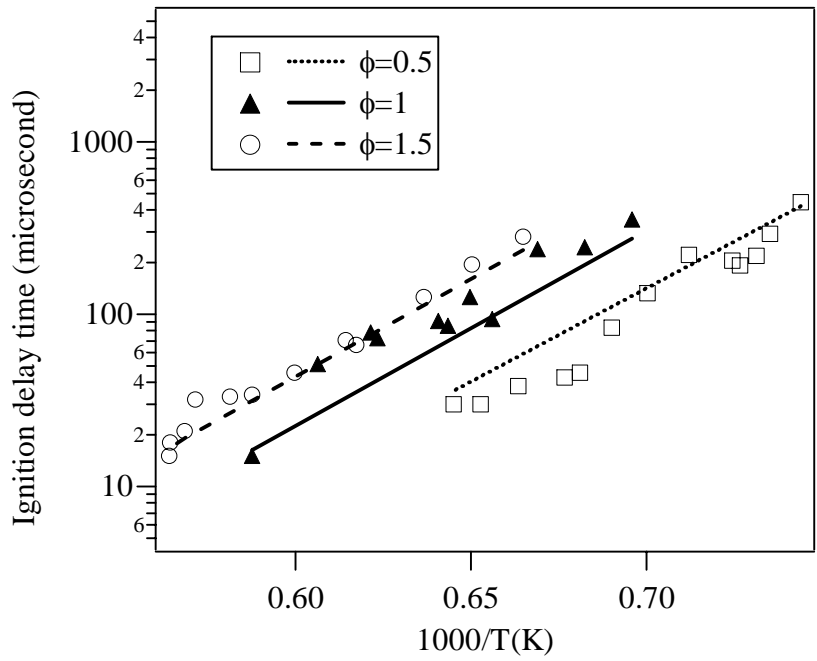

**FIG 3**

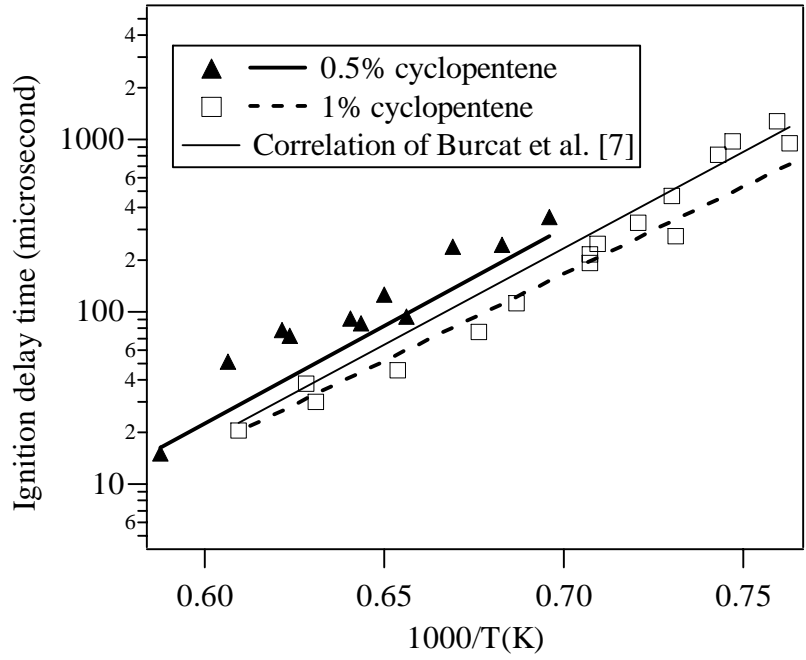

**FIG 4**

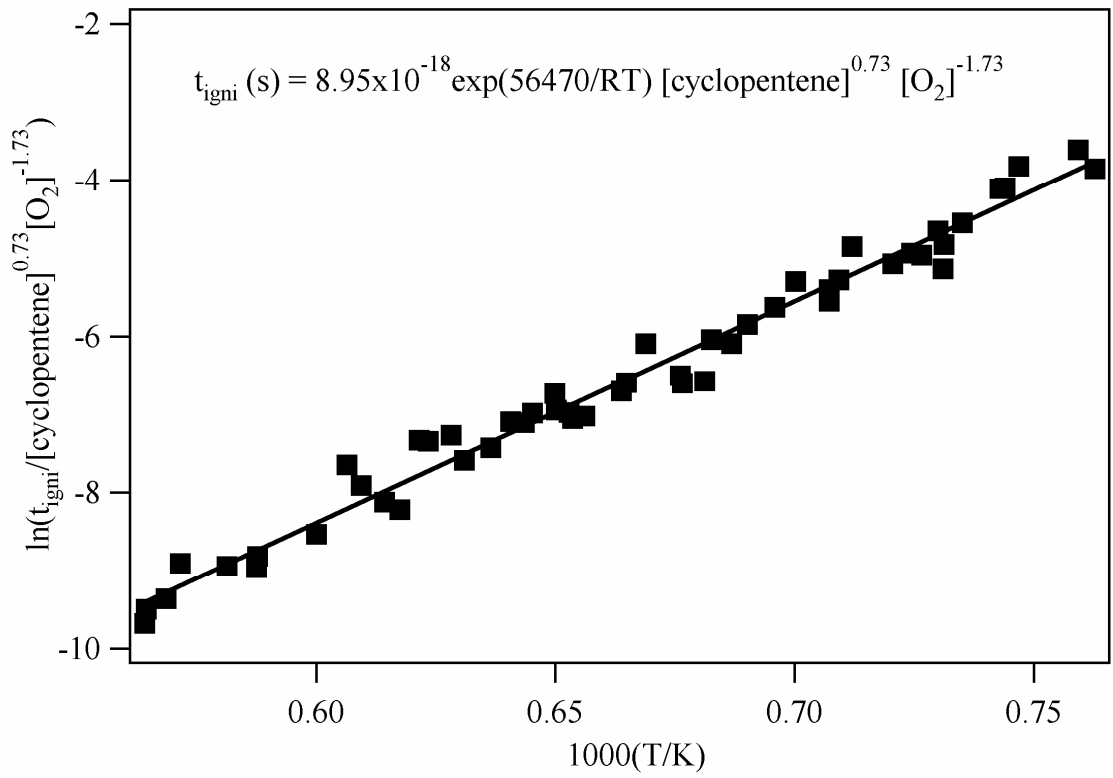

**FIG 5**

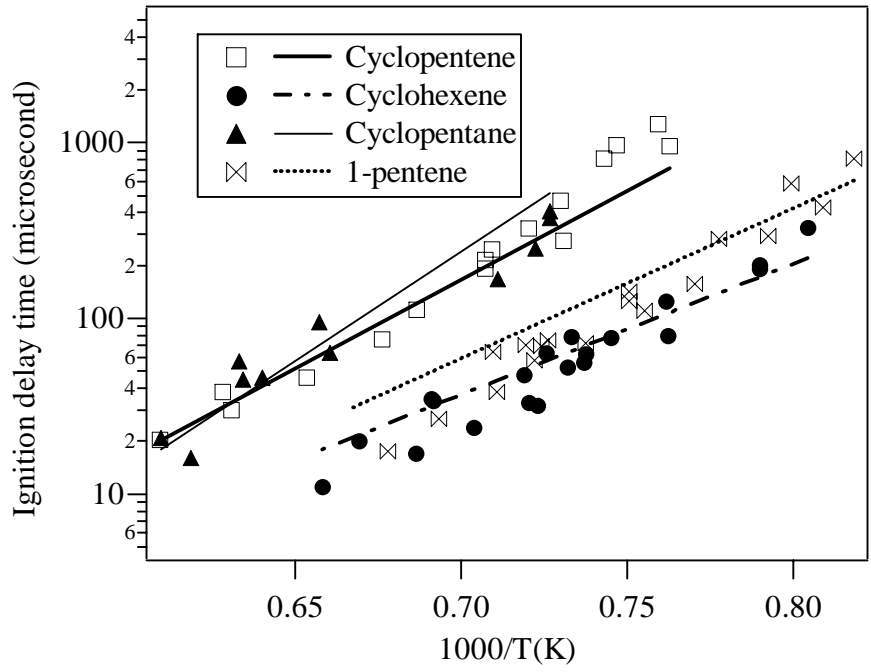

**FIG 6**

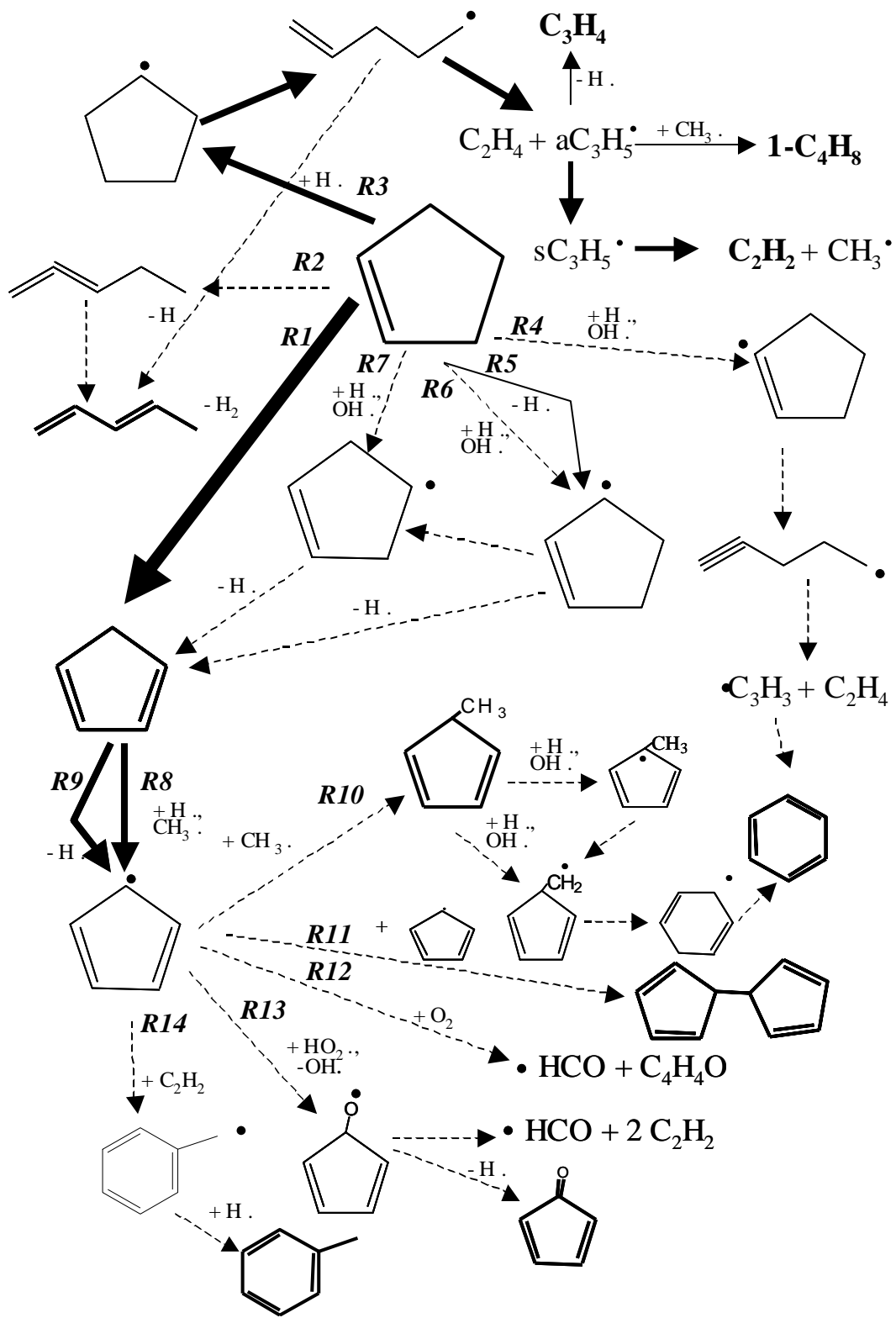

**FIG 7**

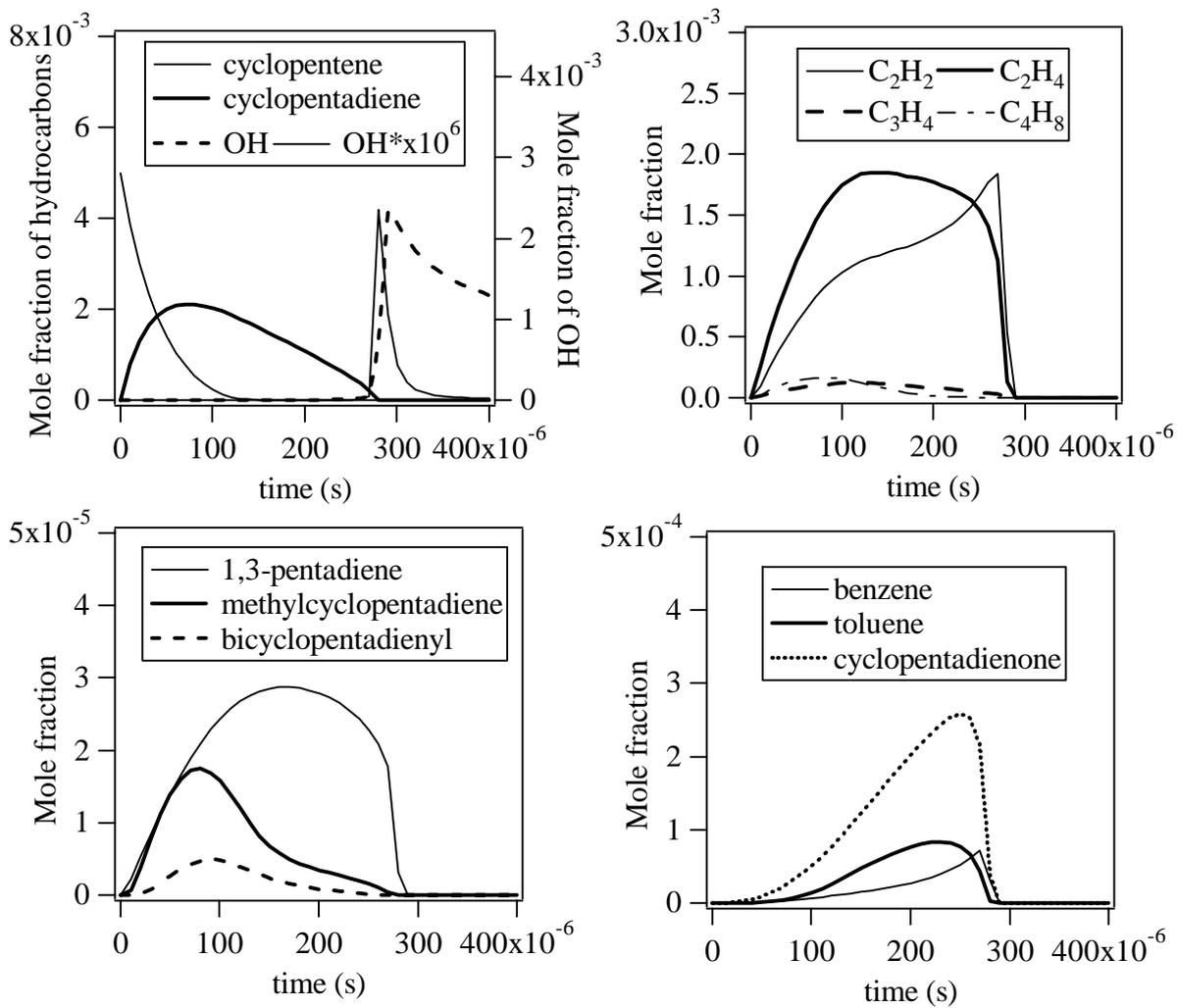

**FIG 8**

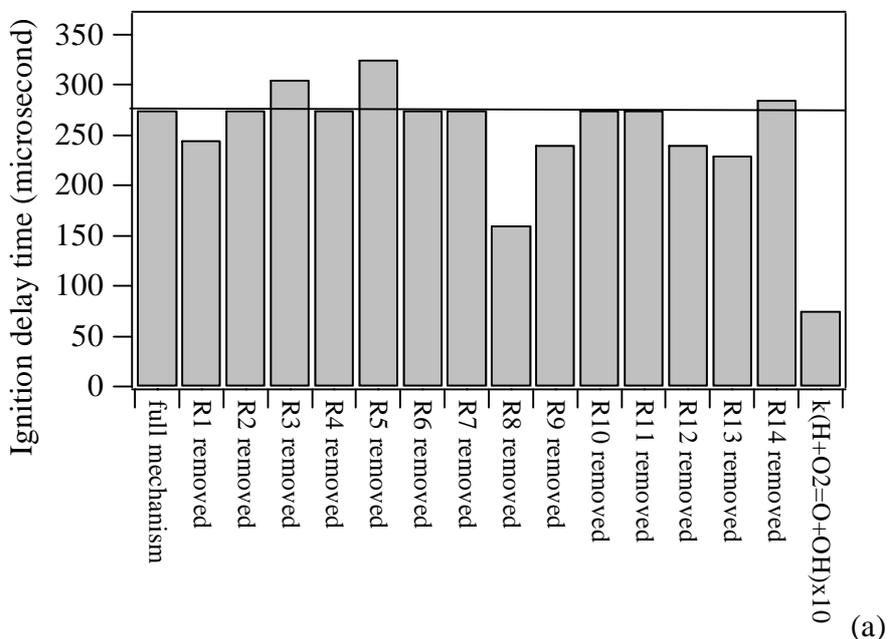

(a)

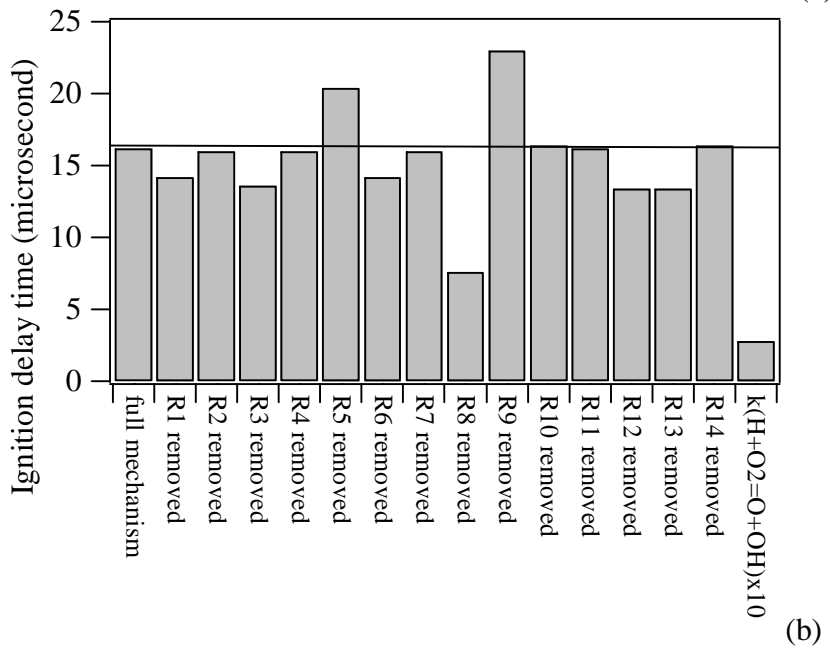

(b)